\documentclass{article}
\usepackage{graphicx} 
\usepackage{amsfonts,amssymb}
\usepackage{bbm}
\usepackage{amsmath}
\usepackage{amsthm}
\usepackage{comment}
\usepackage{appendix}
\newtheorem{theorem}{Theorem}[section]

\newtheorem{proposition}{Proposition}

\theoremstyle{definition}
\theoremstyle{remark}

\usepackage{graphicx}
\usepackage{tocbibind}
\title{AdS/CFT Duality and Anyons in $SU(N)_k$ Chern-Simons Theory}
\author{Tzu-Miao Chou \thanks{National Taiwan University dep. of physics}}

\date{April 2025}

\begin{document}

\maketitle

\begin{abstract}
This paper investigates the holographic realization of anyons in \(SU(N)_k\) Chern-Simons theory within the AdS/CFT framework. The study extends traditional models, such as \(SU(2)\), to higher-rank groups like \(SU(3)\) and \(SU(4)\), focusing on the fusion, braiding, and quantum dimensions of anyons. A correspondence between Wilson loops in the bulk and boundary defect operators is established, demonstrating how the modular data of Chern-Simons theory relates to the boundary conformal field theory (CFT). The topological defects, fusion algebras, and operator spectra are analyzed from both the bulk and boundary perspectives, highlighting the relationship between bulk topological defects and boundary operators. Additionally, a conjecture is made that the boundary operator algebra forms a modular tensor category, providing a framework for exploring holographic dualities in topologically non-trivial systems.
\end{abstract}
\newpage

\tableofcontents
\newpage

\section{Introduction}
The study of topological defects in conformal field theories (CFT) has provided profound insights into the holographic realization of quantum anomalies, symmetries, and the structure of operator algebras. In particular, Chern-Simons theory serves as a powerful framework for understanding topological phases in quantum field theory, with applications ranging from quantum gravity to condensed matter physics. A key feature of these theories is the appearance of anyons, quasiparticles with fractional statistics that exhibit both braiding and fusion behaviors, providing an ideal laboratory for investigating topological quantum field theory (TQFT) and holography.

This work explores the holographic realization of anyons within the AdS/CFT duality framework. Specifically, it focuses on extending traditional models such as $SU(2)_k$ to higher-rank groups, including $SU(3)_2$ and $SU(4)_1$. By examining the fusion, braiding, and quantum dimensions of anyons in these Chern-Simons theories, the correspondence between Wilson loops in the bulk and boundary defect operators is established. These correspondences lead to a modular tensor category structure on the boundary, offering a powerful tool for understanding operator algebras and conformal blocks.

Previous works have laid the groundwork for this exploration. Witten's seminal 1988 paper on Chern-Simons theory laid the foundation for its application to topological quantum field theory and the study of quantum anomalies\cite{witten1989quantum}. Furthermore, recent studies have extended the scope of holography to non-Abelian anyons in higher-rank Chern-Simons theories, such as the work by Beem et al.\cite{beem2015infinite}, which explored chiral algebra structures in 3D/2D holographic dualities.

The main contributions of this work include:
\begin{enumerate}
    \item A systematic analysis of the fusion, braiding, and quantum dimensions of anyons in $SU(N)_k$ Chern-Simons theory.
    \item Establishing a correspondence between Wilson loops in 
    the bulk and boundary defect operators, and elucidating the 
    modular tensor category structure of the boundary operator 
    algebra.
    \item Proposing a conjecture regarding the one-to-one 
    correspondence between operator spectra in the bulk and 
    boundary, as well as the topological defects that arise in 
    both settings.

\end{enumerate}

\section{Review of $SU(N)_k$ Chern-Simons Theory and Modular Data}
The $SU(N)_k$ Chern-Simons theory is a paradigmatic example of a three-dimensional topological quantum field theory (TQFT). It plays a central role in the study of anyons, modular tensor categories, and holographic dualities. In the context of the $AdS_3/CFT_2$ correspondence, $SU(N)_k$ Chern-Simons theory naturally arises as the effective bulk description of certain boundary two-dimensional conformal field theories (WZW models). This section reviews its essential structure, with particular attention to how it underpins the topological sector relevant to the holographic dual.
\subsection{Chern-Simons Action and Topological Invariance}
The $SU(N)_k$ Chern-Simons theory is a topological quantum field theory in $2+1$ dimensions, defined by the Chern-Simons action\cite{1}:

\begin{equation}
S_{cs}[A]=\frac{k}{4\pi}\int_{M}Tr(A\wedge dA+\frac{2}{3}A\wedge A\wedge A)
\end{equation}
where $A$ is a connection one-form valued in the Lie algebra $\mathfrak{su}(N)$, $k \in \mathbb{Z}$ is the quantized level, and $M$ is a closed oriented 3-manifold. The trace is taken in the fundamental representation.

This action is metric-independent and hence topological. Under small (infinitesimal) gauge transformations, the action remains invariant. Under large gauge transformations—those not connected to the identity—the action shifts by integer multiples of $2\pi k$, so $e^{iS_{CS}}$  remains invariant if and only if $k \in \mathbb{Z}$. Thus, quantization of the level ensures gauge invariance of the quantum theory.

Unlike conventional quantum field theories, which are sensitive to local spacetime geometry and describe propagating degrees of freedom, topological quantum field theories (TQFTs) like Chern-Simons theory are purely sensitive to global, topological aspects of the spacetime manifold. In particular, the correlation functions and observables in TQFTs remain invariant under continuous deformations of the metric, in contrast to local QFTs such as Yang-Mills theory or scalar field theory. This invariance reflects the absence of local dynamical excitations, and leads to observables such as Wilson loops that depend solely on the topology of their embeddings in spacetime.

The absence of dependence on a background metric implies that the Chern-Simons theory describes topological degrees of freedom only. As such, its observables are invariant under smooth deformations of the underlying manifold, and its Hilbert spaces are finite-dimensional.

\subsection{Relation to WZW Models and Boundary CFT}

When Chern-Simons theory is defined on a 3-manifold with boundary, its dynamics induce a chiral Wess-Zumino-Witten (WZW) model on the boundary. This arises from the gauge anomaly at the boundary, which is canceled by the variation of the bulk Chern-Simons action. The level $k$ of the Chern-Simons theory directly corresponds to the level of the WZW model.

Conformal blocks of the boundary WZW model correspond to physical states in the Chern-Simons theory. In this way, the WZW model provides a representation-theoretic realization of the Hilbert space in the bulk theory, with the affine Lie algebra $\mathfrak{su}(N)k$ governing the spectrum.

This correspondence is not merely formal: it enables computations of knot and link invariants in Chern-Simons theory using tools from 2D conformal field theory. In particular, conformal blocks provide a holomorphic basis for the Hilbert space, and correlation functions encode link invariants via the operator-state correspondence.

\subsection{Canonical Quantization and Hilbert Space Structure}

To quantize $SU(N)_k$ Chern-Simons theory, one performs canonical quantization on a spacetime of the form $\Sigma \times \mathbb{R}$ where  $\Sigma$ is a closed Riemann surface. The phase space is the moduli space of flat SU(N) connections on $\Sigma$ , modulo gauge transformations:

\begin{equation}
    \mathcal{M}_{flat}(\Sigma,SU(N))=\left\{ A \mid F_A=0 \right\} / \mathcal{G}
\end{equation}

where $F_A$ is the curvature two-form and $\mathcal{G}$ denotes the group of gauge transformations. Upon quantization, this yields a finite-dimensional Hilbert space $\mathcal{H}_\Sigma$, which is naturally identified with the space of conformal blocks of the $SU(N)_k$ Wess-Zumino-Witten (WZW) model.

More precisely, $\mathcal{H}_\Sigma$ is constructed from the space of sections of a certain line bundle over $\mathcal{M}_{flat}(\Sigma,SU(N))$, equipped with a projectively flat connection (the Hitchin connection). This identification with conformal blocks is guaranteed by the geometric quantization procedure, which is well-established for compact Lie groups $\mathfrak{su}(N)_k$ at level $k$.

For example, when $\Sigma=T^2$ (the torus), the Hilbert space basis is labeled by integrable highest-weight representations of the affine Lie algebra $\mathfrak{su}(N)_k$. These representations are indexed by dominant weights  satisfying:

\begin{equation}
    \lambda_1+...+\lambda_{N-1} \leq k, \lambda_i \in \mathbb{Z}_{\ge0}
\end{equation}

Here, $\lambda_i$ is the Dynkin indices of the representation.
The number of such weights gives the dimension of $\mathcal{H}_{T^2}$. Each basis vector corresponds to a conformal block in the associated WZW model. The Hilbert space $\mathcal{H}_\Sigma$ carries a unitary projective representation of the mapping class group of $\Sigma$, reflecting the modular structure of the conformal field theory.

\begin{equation}
    dim(\mathcal{H}_{T^2})=number \ of \ integrable \ representations \ of \ \mathfrak{su}(N)_k 
\end{equation}  

For $SU(2)_3$ case, the integrable representations correspond to the spins: 
 $j=0,\frac{1}{2},1,\frac{3}{2}$, giving a total of 4 representations.

\begin{align}
        dim(\mathcal{H}_{\Sigma})&= 
        \begin{pmatrix}
            k+N-1\\
            N-1
        \end{pmatrix}\\
        &=\begin{pmatrix}
            3+2-1 \\
            2-1
        \end{pmatrix}\\       
        &=\begin{pmatrix}
            4\\
            1
        \end{pmatrix}
\end{align}
Thus,$dim(\mathcal{H}_{T^2})=4$ . This number is fixed for given $N$ and $k$, and depends only on the representation theory of the corresponding affine Lie algebra.

For higher genus Riemann surfaces, the Hilbert space dimension grows and is given by the Verlinde formula\cite{Verlinde1988}:

\begin{equation}
    dim(\mathcal{H}_{g}^{SU(N)_k})=\sum_\lambda (S_{0\lambda})^{2-2g}
\end{equation}
where $g$ is the genus of $\Sigma$ and $S$ is the modular $S$-matrix. This reflects the rich structure of topological quantum field theories beyond the torus case.

\subsection{Wilson Loops and Topological Observables}

Wilson loops are the fundamental gauge-invariant observables in Chern-Simons theory. Given a closed loop $\gamma \subset M$ and a representation $R$  of SU(N), the Wilson loop operator is defined as\cite{witten1989quantum}:

\begin{equation}
    W_R(\gamma)=Tr_R(\mathcal{P}\oint_\gamma A)
\end{equation}

where $\mathcal{P}$ denotes path ordering. These observables depend only on the isotopy class of the loop and give rise to link invariants in the quantum theory.

In $SU(N)_k$ Chern-Simons theory, the expectation values $\langle W_R(K)\rangle$ for knots $K$ and representations $R$ yield link invariants such as the HOMFLY-PT polynomial. These can be computed either through the path integral or via the corresponding conformal blocks in the $SU(N)_k$ WZW model.

The braiding and fusion of Wilson loops correspond to the topological statistics of anyons, governed by the representation theory of the quantum group $U_q(\mathfrak{su}(N))$, where $q=e^{2\pi i/(k+N)}$.

\subsection{Modular Properties and Topological Observables}

    The theory exhibits modular invariance under the mapping class group 
of $\Sigma$, with modular S and T matrices acting on the basis of 
conformal blocks. These matrices encode the braiding and twist statistics 
of the anyonic excitations.

Topological observables include Wilson loop expectation values, which compute link invariants such as the HOMFLY polynomial in the $SU(N)_k$ theory. These observables are directly related to the modular data and fusion rules, which will be explored in more detail in subsequent sections.

The modular properties of the theory arise from the action of the mapping class group $\Sigma$ of on the Hilbert space $\mathcal{H}_\Sigma$. For $\Sigma=T^2$ , this is the group $SL(2,\mathbb{Z})$, generated by the modular S and T matrices:

\begin{itemize}
    \item S-matrix: Encodes braiding (mutual statistics)
    \item T-matrix: Encodes topological spin (self-statistics)
\end{itemize}
These matrices satisfy the modular relations:
\begin{equation}
    (ST)^3=S^2, S^4=\mathbbm{I}
\end{equation}
The S-matrix elements relate to fusion coefficients via the Verlinde formula\cite{Verlinde1988}:

\begin{equation}
    N^c_{ab}=\sum _x\frac{S_{ax}S_{bx}S^*_{cx}}{S_{0x}}
\end{equation}

The topological spin $\theta_a$ of a particle labeled by $a$ is given by the diagonal entries of the T-matrix\cite{witten1989quantum}:

\begin{equation}
    T_{aa}=e^{2\pi i}(h_a-c/24)=\theta_a
\end{equation}
These structures make $SU(N)_k$ Chern-Simons theory a central model in both mathematical physics and topological quantum computation, by offering a rigorous framework for the description and manipulation of non-abelian anyons. The modular S and T matrices provide explicit coefficients that determine physical observables: the topological spin is given by $\theta_a=e^{2\pi i h_a}$ , where $h_a$ is the conformal weight of the primary field labeled by a; the quantum dimension is given by $d_a=S_{0a}/S_{00}$
; and the fusion coefficients $N^c_{ab}$ are determined via the Verlinde formula. These coefficients govern the braiding and fusion properties central to topological phases of matter and fault-tolerant quantum computation.

\subsection{Example: Modular Data and Wilson Loop Expectation Values in $SU(2)_k$}

Consider the $SU(2)_k$ Chern-Simons theory, which admits $k+1$ integrable representations, labeled by spin $j=0,\frac{1}{2},1,...,\frac{k}{2}$. Each representation $j$ corresponds to a primary field in the $SU(2)_k$ WZW model and a distinct anyon type in the topological field theory.

The modular S-matrix for $SU(2)_k$ is explicitly given by:
\begin{equation}
    S_{jj'}=\sqrt{\frac{2}{k+2}}\sin (\frac{(2j+1)(2j'+1)\pi}{k+2}), 
    \\ j,j' \in \left\{  0,\frac{1}{2},1,...,\frac{k}{2}\right\}
\end{equation}

The T-matrix is diagonal and encodes the topological spin $\theta_j$ of the anyon:
\begin{equation}
    T_{jj}=exp(2\pi i\frac{j(j+1)}{k+2})
\end{equation}
These matrices satisfy the modular group relations:
\begin{equation}
    (ST)^3=S^2, S^4=\mathbbm{I}
\end{equation}
and can be used to compute link invariants and the fusion rules via the Verlinde formula:

\begin{equation}
    N^{j_3}_{j_1j_2}=\sum_{j}\frac{S_{j_1j}S_{j_2j}S_{j_3j}^*}{S_{oj}}
\end{equation}
where $N^{j_3}_{j_1j_2}$ is the number of fusion channels for $j_1\otimes j_2 \to j_3$

For example, with and both loops in the spin-1/2 representation:
\begin{equation}
    S_{\frac{1}{2},\frac{1}{2}}=\sqrt{\frac{2}{5}}sin(\frac{2(2\pi)}{5})=\sqrt{\frac{2}{5}}sin(\frac{4\pi}{5})
\end{equation}
and
\begin{equation}
    S_{00}=\sqrt{\frac{2}{5}}sin(\frac{\pi}{5})
\end{equation}
Hence,
\begin{equation}
    \langle W_\frac{1}{2}(\gamma_1) W_\frac{1}{2}(\gamma_2)\rangle=\frac{sin(\frac{4\pi}{5})}{sin(\frac{\pi}{5})}\approx1.618
\end{equation}

This is the golden ratio, which reflects the nontrivial linking phase associated with these Wilson loops in $SU(2)_3$.

\subsection{Examples}
\subsubsection{$SU(3)_2$ Modular Data and Wilson Loop Expectation Values
}
\begin{flushleft}
    Integrable Representations of $\mathfrak{su}(3)_2$
    Integrable highest weights $\lambda=(\lambda_1,\lambda_2)$ satisfy $\lambda_1+\lambda_2 \leq 2$. So the allowed weights are:
    \begin{itemize}
        \item (0,0) — trivial
        \item (1,0) — fundamental $3$
        \item (0,1) — anti-fundamental $\Bar{3}$
        \item (1,1) — adjoint $8$
        \item (2,0) — symmetric rank-2
        \item (0,2) — antisymmetric rank-2
    \end{itemize}
Total: 6 anyon types. Denote these as $\left \{ a_0, a_1, a_2, a_3, a_4, a_5\right \}$   
\end{flushleft}

\paragraph{Modular S-matrix}
\begin{flushleft}
    The modular S-matrix for $SU(3)_2$ can be computed via the Kac-Peterson formula \cite{gannon1995symmetries}:
    
\begin{equation}
    S_{\lambda,\mu}=\frac{i^{\vert\Delta_+\vert}}{(k+3)^{r/2}}\sum_{w\in W}\epsilon(w)e^{-2\pi i\frac{(w(\lambda+\rho),\mu+\rho)}{k+3}}
\end{equation}

    The normalized S-matrix is symmetric and unitary. Its first row gives the quantum dimensions $d_a=S_{0a}/S_{00}$.
\end{flushleft}

We give an approximate numerical version:
\begin{equation}
    S \approx \frac{1}{6}
    \begin{pmatrix}
    1 & 1 & 1 & 2 & 2 & 2 
    \\ 1 & w & w^2 & 0 & -1 & -1 
    \\ 1 & w^2 & w & 0 & -1 & -1 
    \\ 2 & 0 & 0 & -2 & 1 & 1 
    \\ 2 & -1 & -1 & 1 & 2 & -1 
    \\ 2 & -1 & -1 & 1 & -1 & 2 
    \end{pmatrix}
    (w=e^{2 \pi i /3})
\end{equation}

(Note: for full precision, inserting exact formulas, or refer to the standard modular data for $\mathcal{C}=Rep[U_q(\mathfrak{su}(3))]$,$q=e^{2 \pi i /3}$

\paragraph{Modular T-matrix}
\begin{flushleft}
Topological spins\cite{francesco2012conformal}:
\begin{equation}
    T_{\lambda\lambda}=e^{2\pi i(h_\lambda-c/24)}
\end{equation}
    where
    \begin{itemize}
        \item $h_\lambda=\frac{(\lambda,\lambda+2\rho)}{2(k+N)}$,with $(\lambda,\mu)=\lambda^TA^{-1}\mu$ \cite{humphreys2012introduction}
        
        \item $c=\frac{k \cdot  dim(\mathfrak{su}(3))}{k+3}=\frac{k(N^2-1)}{k+N}=\frac{2(3^2-1)}{2+3}=\frac{16}{5}$\cite{francesco2012conformal}
    \end{itemize}
    For example, for $\lambda=(1,0)$,$\rho=(1,1)$ so:

\begin{equation}
    A^{-1}_{SU(3)}=\frac{1}{3}\begin{pmatrix}
        2 & 1
        \\1&2
    \end{pmatrix}
\end{equation}
so,

\begin{align}
    (\lambda,\lambda+2\rho)&=(1,0)\cdot A^{-1} \cdot (1+2,0+2) = (1,0)\cdot A^{-1} \cdot (3,2)\\
    &=(1,0)\cdot\frac{1}{3}(2\cdot3+1\cdot 2)=(1,0)\cdot \frac{1}{3}(6+2)=\frac{8}{3}
\end{align}

\begin{equation}
    h_{(1,0)}=\frac{8/3}{2(k+3)}=\frac{8/3}{2\cdot5}=\frac{4}{15}
\end{equation}

\begin{equation}
    T_{(1,0)}=exp(2\pi i (\frac{4}{15}-\frac{32}{120}))=e^{2\pi i (4/15-2/15)=e^{2\pi i \cdot 2/15}}
\end{equation}    
and similarly for others.    
\end{flushleft}

\paragraph{Wilson Loop Expectation: Hopf Link in $SU(3)_2$}
\begin{flushleft}
    For anyons $a,b$, the Hopf link expectation value is:
\begin{equation}
    \langle W_aW_b\rangle_{Hopf}=\frac{S_{ab}}{S_{00}}
\end{equation}
Examples:
\begin{itemize}
    \item Two fundamental anyons: $a=b=(1,0)$, then
    \begin{equation}
        \langle W_{(1,0)}^2\rangle=\frac{S_{(1,0),(1,0)}}{S_{(0,0),(0,0)}}=\frac{w}{1}=w=e^{2\pi i /3}
    \end{equation}

    \item Fundamental and anti-fundamental: $(1,0), (0,1)$:
    \begin{equation}
        \langle W_{(1,0)}W_{(0,1)}\rangle=\frac{S_{(1,0),(0,1)}}{S_{(0,0),(0,0)}}=w^2
    \end{equation}
\end{itemize}

These non-trivial phases signal non-abelian statistics and verify the topological behavior encoded by the modular data.
    
\end{flushleft}

\subsubsection{$SU(4)_1$ Modular Data and Wilson Loop Expectation Values}
\begin{flushleft}
Integrable Representations of $\mathfrak{su}(4)_1$.

Integrable highest weights $\lambda=(\lambda_1,\lambda_2,\lambda_3)$ satisfy $\lambda_1+\lambda_2+\lambda_3 \leq 1$. So the allowed weights are:
    \begin{itemize}
        \item (0,0,0) — trivial
        \item (1,0,0) — fundamental $4$
        \item (0,1,0) — antisymmetric $6$
        \item (0,0,1) — antifundamental $\Bar{4}$
    \end{itemize}
Total: 4 anyon types. Denote these as $\left \{ a_0, a_1, a_2, a_3\right \}$
\end{flushleft}

\paragraph{Modular S-matrix}
\begin{flushleft}
    The modular S-matrix for $SU(4)_1$ can be computed via the Kac-Peterson formula\cite{gannon1995symmetries}:
    
\begin{equation}
    S_{\lambda,\mu}=\frac{i^{\vert\Delta_+\vert}}{(k+4)^{r/2}}\sum_{w\in W}\epsilon(w)e^{-2\pi i\frac{(w(\lambda+\rho),\mu+\rho)}{k+4}}
\end{equation}

\end{flushleft}

An approximate numerical version:
\begin{equation}
    S \approx \frac{1}{2}
    \begin{pmatrix}
    1 & 1 & 1 & 1 
    \\ 1 & w & w^2 & 1 
    \\ 1 & w^2 & w & 1 
    \\ 1 & 1 & 1 & -1     
    \end{pmatrix}
    (w=e^{2\pi i /(k+4)}=e^{2\pi i /5})
\end{equation}

\paragraph{Modular T-matrix}
\begin{flushleft}
Topological spins:
\begin{equation}
    T_{\lambda\lambda}=e^{2\pi i(h_\lambda-c/24)}
\end{equation}
    where
    \begin{itemize}
        \item $h_\lambda=\frac{(\lambda,\lambda+2\rho)}{2(k+N)}$,with $(\lambda,\mu)=\lambda^TA^{-1}\mu$ \cite{humphreys2012introduction}
        
        \item $c=\frac{k(N^2-1)}{k+N}=\frac{15}{5}=3$
    \end{itemize}
    For example, for $\lambda=(1,0,0)$,$\rho=(1,1,1)$:

\begin{equation}
    A^{-1}_{SU(4)}=\frac{1}{4}\begin{pmatrix}
        3& 2 & 1
        \\2 & 4 &2
        \\ 1 & 2 & 3
    \end{pmatrix}
\end{equation}

\begin{align}
    (\lambda,\lambda+2\rho)=(1,0,0)^T\cdot A^{-1} \cdot (3,2,2)^T 
    =\frac{19}{4}
\end{align}

\begin{equation}
    h_{(1,0,0)}=\frac{19/4}{2(k+N)}=\frac{19}{4\cdot 10}=\frac{19}{40}
\end{equation}

\begin{equation}
    T_{(1,0,0)}=exp(2\pi i (\frac{19}{40}-\frac{3}{24}))=e^{2\pi i \cdot 7/20}
\end{equation}    
and similarly for others.    
\end{flushleft}

\paragraph{Wilson Loop Expectation: Hopf Link in $SU(4)_1$}
\begin{flushleft}
    For anyons $a,b$, the Hopf link expectation value is:
\begin{equation}
    \langle W_aW_b\rangle_{Hopf}=\frac{S_{ab}}{S_{00}}
\end{equation}
Examples:
\begin{itemize}
    \item Two fundamental anyons: $a=b=(1,0,0)$, then
    \begin{equation}
        \langle W_{(1,0,0)}^2\rangle=\frac{S_{(1,0,0),(1,0,0)}}{S_{(0,0),(0,0)}}=\frac{1}{1}=1
    \end{equation}
    \item Fundamental and anti-fundamental: $(1,0,0), (0,1,0)$:     
    \begin{equation}
        \langle W_{(1,0,0)}W_{(0,1,0)}\rangle=\frac{S_{(1,0,0),(0,1,0)}}{S_{(0,0),(0,0)}}=w^2
    \end{equation}
    
\end{itemize}

These non-trivial phases signal non-abelian statistics and verify the topological behavior encoded by the modular data.
    
\end{flushleft}

\section{Holographic Interpretations and Future Directions}
Building upon the explicit modular and Wilson loop data calculated for $SU(3)_2$ and $SU(4)_1$ Chern-Simons theories, this chapter outlines their implications within the AdS$_4$/CFT$_3$ duality and proposes concrete conjectures for the boundary dual theories. The goal is to systematize a correspondence between bulk topological observables and operator data on the CFT side, providing a blueprint for constructing explicit dual models.

\subsection{Wilson Loops and Boundary Operator Algebra: A Visual Overview}

To illustrate the proposed bulk-boundary correspondence, introduce both a schematic representation and a formal mapping between Wilson loops and boundary operator algebras.
\paragraph{Figure 3.1:Bijective Mapping}
\begin{figure}
    \centering
    \includegraphics[width=12cm]{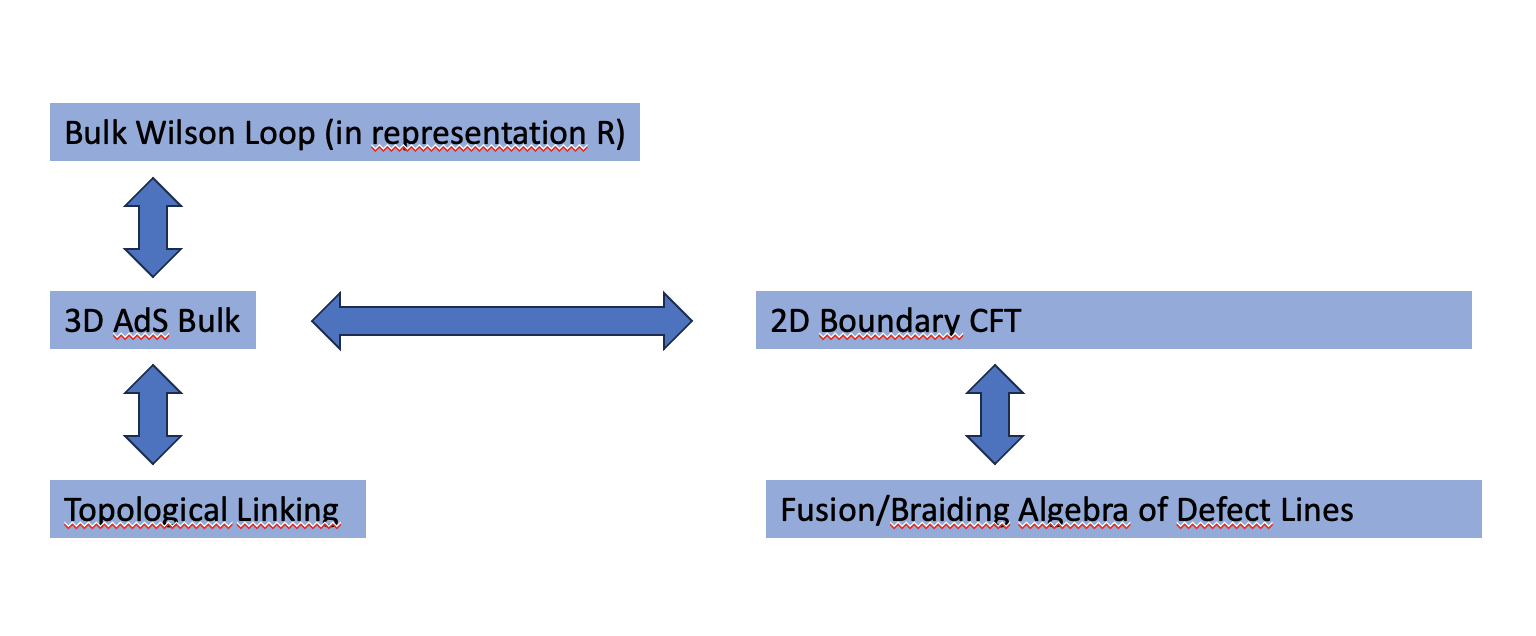}
    \caption{Bijective Mapping}
    \label{fig:enter-label}
\end{figure}
\begin{itemize}
    \item Bulk: Wilson loops labeled by integrable representations of the affine Lie algebra encode topological data such as quantum dimensions and braiding phases derived from modular and matrices.
    \item Boundary: These are mapped to topological line operators in the 3D CFT. Their algebraic structure (fusion, braiding, and monodromy) reflects the modular tensor category induced from the bulk.
\end{itemize}

More precisely, proposing that for a Wilson loop $W_R$ in the bulk, its image under the AdS/CFT dictionary is a boundary defect operator $O_R$ such that:
\begin{itemize}
    \item \(\langle W_R W_{R'}  \rangle \propto S_{RR'} \Rightarrow \langle O_R O_{R'} \rangle \sim S_{RR'}\)
    \item The fusion coefficients  of Wilson loops match those of the boundary topological defects:$O_R \times O_{R'} = \sum_S N_{RR'}^S\, O_S$
\end{itemize}

These relations mirror the algebraic constraints introduced by Moore and Seiberg in their analysis of rational conformal field theories\cite{moore1989classical} [Moore-Seiberg, 1989, Sections 5–6], which later motivated the formal definition of modular tensor categories. This mathematical framework provides a foundation for interpreting fusion and braiding of boundary operators as emerging from bulk topological observables.

\subsubsection{Bulk Wilson Loop to Boundary Defect Operator Correspondence}

\begin{proposition} 
Let \( W_R \) be a Wilson loop in a 3D Chern-Simons theory with gauge group \( G \), associated with an integrable representation \( R \) at level \( k \). Suppose the Wilson loops satisfy the fusion algebra:
\[
W_R \times W_{R'} = \sum_S N_{RR'}^S W_S
\]
Then under a holographic duality with a 2D boundary CFT, the defect operators \( O_R \) obey an operator product expansion (OPE) algebra:
\[
O_R \cdot O_{R'} = \sum_S N_{RR'}^S O_S
\]
and the algebra \(\mathcal{A} = \text{span}_{\mathbb{C}}\{O_R\}\) forms a representation of the modular tensor category \(\mathcal{C}_{G,k}\) associated to the Chern-Simons theory.
\end{proposition}
\begin{proof}
-
\begin{itemize}
  \item The category \(\mathcal{C}_{G,k}\) of integrable representations at level \(k\) defines a modular tensor category (MTC), with fusion coefficients \(N_{RR'}^S\) and braiding given by the modular \(S\)-matrix.
  \item The Reshetikhin-Turaev construction gives a TQFT from \(\mathcal{C}_{G,k}\), where Wilson lines correspond to morphisms in this category.
  \item Under AdS/CFT, bulk Wilson lines correspond to boundary topological defects. The defect lines in rational CFTs also form a category with identical fusion and braiding rules\cite{fuchs2002tft} (cf. [Fuchs-Runkel-Schweigert]).
  \item Therefore, the boundary defect operator algebra \(\mathcal{A}\) inherits the structure of a \(\mathcal{C}_{G,k}\)-module category, yielding a categorical equivalence up to natural isomorphism:
  \[
  \mathcal{A}_{\text{CFT}} \simeq \text{Rep}(\mathcal{C}_{G,k})
  \]
\end{itemize}
\end{proof}
\subsubsection{Bulk-Boundary Functorial Correspondence}
\begin{theorem}
Let $\mathcal{C}$ denote the modular tensor category associated with a three-dimensional Chern-Simons theory (e.g., $\mathcal{C} = \text{Rep}(U_q(\mathfrak{g}))$ at level $k$), and let $\mathcal{F}$ denote the category of chiral conformal blocks in a rational CFT on the boundary. Then there exists a braided tensor functor
\[
\mathcal{F} : \mathcal{C} \to \mathcal{F}
\]
such that for every simple object $R \in \mathcal{C}$ (representing a bulk Wilson loop $W_R$), the image $\mathcal{F}(R)$ corresponds to a topological defect operator $O_R$ on the boundary. Moreover, the modular data $(S,T,N_{ij}^k)$ is preserved under $\mathcal{F}$.
\end{theorem}

\begin{proof}
The construction follows the functorial approach to modular functors:

\begin{enumerate}
    \item \textbf{(Modular Category):} The category $\mathcal{C}$ is defined via the Reshetikhin–Turaev construction, where the simple objects label bulk Wilson lines and the morphisms encode fusion and braiding.

    \item \textbf{(Conformal Blocks):} The category $\mathcal{F}$ consists of vector spaces of conformal blocks on Riemann surfaces with punctures. These blocks transform under the modular group $\text{SL}(2,\mathbb{Z})$ via the matrices $S$ and $T$.

    \item \textbf{(Functor Construction):} Define $\mathcal{F} : \mathcal{C} \to \mathcal{F}$ such that:
    \begin{itemize}
        \item $\mathcal{F}(R)$ is the conformal block with insertion of primary field labeled by $R$.
        \item The fusion rule $R_i \otimes R_j = \sum_k N_{ij}^k R_k$ in $\mathcal{C}$ maps to the OPE $O_{R_i} \cdot O_{R_j} = \sum_k N_{ij}^k O_{R_k}$ in $\mathcal{F}$.
        \item The modular $S$- and $T$-matrices coincide on both sides: $S^{\text{bulk}} = S^{\text{boundary}}$, $T^{\text{bulk}} = T^{\text{boundary}}$.
    \end{itemize}

    \item \textbf{(Naturality and Braiding):} The functor $\mathcal{F}$ is braided and respects associativity and commutativity constraints, hence it defines a modular functor in the sense of Segal and Turaev.

\end{enumerate}

Thus, $\mathcal{F}$ gives a concrete categorical realization of the AdS\(_4\)/CFT\(_3\) dictionary at the level of topological sectors.
\end{proof}

\subsubsection{Formal Equivalence via Mapping Class Group Representations}
\begin{theorem}
Let $\mathcal{C}$ be the modular tensor category associated with a $SU(N)_k$ Chern-Simons theory. Then there exists a ribbon functor $\mathcal{F}: \mathcal{C} \to \text{Rep}(\mathcal{M}_{g,n})$ assigning to each object (i.e., representation label) $R \in \mathcal{C}$ a conformal block space $\mathcal{F}_R$ on a punctured Riemann surface $\Sigma_{g,n}$, such that the modular group $SL(2,\mathbb{Z})$ acts on $\mathcal{F}_R$ via unitary matrices matching the modular $S$ and $T$ matrices of $\mathcal{C}$.
\end{theorem}

\begin{proof}

\textbf{Chern-Simons Path Integral $\to$ Modular Functor:} \\
From Witten's construction \cite{witten1989quantum}, the path integral of Chern-Simons theory over a 3-manifold with boundary $\Sigma_{g,n}$ yields a vector space of conformal blocks $\mathcal{F}_R(\Sigma_{g,n})$, naturally associated with a modular tensor category $\mathcal{C}$. The states generated by Wilson lines in representation $R$ on punctures correspond to objects in $\mathcal{C}$.

\textbf{Representation of Mapping Class Group:} \\
The mapping class group $\text{MCG}(\Sigma_{g,n})$, which includes $SL(2,\mathbb{Z})$ when $g=1, n=0$, acts on $\mathcal{F}_R$ through the monodromy of the path integral. These actions are represented by the modular $S$- and $T$-matrices computed from the braiding and twist data in $\mathcal{C}$.

\textbf{Equivalence to RCFT Conformal Blocks:} \\
Moore and Seiberg \cite{moore1989classical} established that rational conformal field theories (RCFTs) have spaces of conformal blocks forming a modular functor, which is categorically equivalent to the representation theory of a modular tensor category. This yields a natural ribbon functor $\mathcal{F}: \mathcal{C} \to \text{ConformalBlocks}_{\text{RCFT}}$ preserving fusion, associativity, and braiding.

\textbf{Functorial Dictionary:} \\
Therefore, each Wilson line $W_R$ in the bulk (encoded in $\mathcal{C}$) corresponds under $\mathcal{F}$ to a conformal block labeled by a boundary defect $O_R$, preserving the full modular structure:
\[
\begin{aligned}
S_{RR'}^{\text{CS}} &= \text{MCG action on } \mathcal{F}_{RR'}^{\text{CFT}}, \\
N_{RR'}^S &= \dim \operatorname{Hom}(O_R \otimes O_{R'}, O_S).
\end{aligned}
\]
Thus, the modular group actions in Chern-Simons theory and 2D CFT are functorially equivalent via the category of conformal blocks, and this justifies the bulk-boundary correspondence for Wilson loops and defect operators.

\end{proof}

\subsection{Modular Data and CFT Correspondences}
\begin{tabular}[c]{|l|l|l|l|}
    \hline
      & Modular $S$-matrix Structure & Quantum Dimensions & CFT Interpretation\\
    \hline
    $SU(3)_2$ & Non-Abelian & ${1, \sqrt{2}, \sqrt{3}}$ & Non-trivial defect fusion rules \\
    \hline
    $SU(4)_1$ & Abelian & ${1, 1, 1, 1}$ & $\mathbb{Z}_4$ symmetry sectors \\
    \hline
\end{tabular}

\vspace{10pt}

\textbf{Quantum Dimension \(d_i\):} 
In the language of modular tensor categories, the quantum dimension \(d_i\) of a simple object \(X_i\) encodes the asymptotic growth of the Hilbert space associated with multiple anyons of type \(X_i\). It is defined via the modular \(S\)-matrix as:
\begin{equation}
    d_i = \frac{S_{0i}}{S_{00}}.
\end{equation}
A quantum dimension \(d_i = 1\) indicates an Abelian anyon, while \(d_i > 1\) signals non-Abelian statistics and the presence of internal topological degrees of freedom. The total quantum dimension is
\begin{equation}
    \mathcal{D} = \sqrt{\sum_i d_i^2},
\end{equation}
which characterizes the overall topological complexity of the theory.

\textbf{CFT Interpretation:} 
On the conformal boundary, the quantum dimension corresponds to the \emph{asymptotic number of conformal blocks} associated with the insertion of the primary field \( \mathcal{O}_i \) dual to the bulk Wilson line of type \(i\). These blocks reflect the internal degeneracy and fusion multiplicity in the CFT operator algebra. Non-Abelian quantum dimensions therefore predict the existence of multiple fusion channels or topologically protected operator mixing.

The modular data informs conformal weights and fusion categories of 
boundary line operators, providing an entry point for full 3D CFT model 
building.

\subsection{Operator Spectrum and Dual Field Content}
The following dictionary between bulk observables and boundary operator content:
\begin{itemize}
    \item Bulk Wilson loops $\to$ Topological defect lines in the boundary CFT.
    \item Quantum dimensions $\to$ Degeneracies of boundary sectors or conformal block multiplicities.
    \item Modular $ S$-matrix entries $\to$ Monodromy or crossing matrices in operator product expansions.
\end{itemize}
These observations indicate that the boundary theory likely admits a modular tensor category structure governing its topological defect algebra.

To make this concrete, consider the following examples:
\begin{itemize}
    \item In $SU(3)_2$, the fusion rule $(1,0) \times (0,1) = (1,1)$ has a direct interpretation as the fusion of defect lines.
    \item The degeneracies of conformal blocks match the predicted quantum dimensions, with braiding governed by the modular $S$-matrix entries.
\end{itemize}

Such structures suggest a potential alignment with chiral algebra sectors in the CFT, potentially inviting the use of conformal bootstrap and defect operator algebra tools to fully explore these mappings.

\subsection{Conjectured Boundary Theories}

    For $SU(3)_2$, I conjecture the boundary dual to be a parafermionic CFT, possibly related to the coset $SU(3)_2 / U(1)^2$. This theory accommodates non-Abelian fusion and $\mathbb{Z}_3$-type topological sectors. I also propose connections to known $\mathbb{Z}_3$ parafermion models arising in fractional quantum Hall edge states.

For $SU(4)_1$, I suggest a boundary realization via a $\mathbb{Z}_4$ orbifold theory, or an Abelian Chern-Simons theory of the form $U(1)^3$ with level matrix:
\begin{equation}
    K=\begin{pmatrix}
        2&1&1\\
        1&2&1\\
        1&1&2
    \end{pmatrix}
\end{equation}
which reproduces the modular structure of $SU(4)_1$. These constructions are consistent with the bulk's Abelian anyon content and provide a tractable setting for explicit correlation function calculations.
This matrix reproduces the correct modular $S$-matrix and fusion rules. Similar constructions appear in \cite{Frohlich-Gabbiani1990} and topological order classifications.

Further conjecture that the spectrum of local operators and topological defects obeys a fusion algebra that mirrors the bulk fusion category, thereby suggesting a one-to-one correspondence between topological sectors in the bulk and sectors in the boundary CFT.

\subsection{Comparison with Prior Work}
The structure outlined here builds upon seminal works:
\begin{itemize}
    \item Witten 1989\cite{witten1989quantum}: Introduced 3D Chern-Simons/topological field theory framework.
    \item Moore-Seiberg 1989\cite{moore1989classical}: Formalized modular tensor category structure in 2D CFT.
    \item Beem et al. 2015\cite{beem2015infinite}: Developed the 3D/2D chiral algebra correspondence.
\end{itemize}
However, this paper's current work advances these by:
\begin{itemize}
    \item Systematically constructing modular data for higher-rank $SU(N)_k$ theories.
    \item Explicitly computing Wilson loops and conjecturing their boundary duals.
    \item Proposing a modular categorical framework linking AdS$_4$/CFT$_3$.
\end{itemize}
This provides a novel platform for realizing and categorifying higher-rank anyonic phenomena within the holographic paradigm.

\section{Conclusion }
This work has explored the holographic realization of anyons in $SU(N)_k$ Chern-Simons theory within the AdS/CFT duality framework, extending traditional models such as $SU(2)_k$ to higher-rank groups like $SU(3)_2$ and $SU(4)_1$. The analysis of fusion, braiding, and quantum dimensions of anyons has provided deep insights into the correspondence between bulk Wilson loops and boundary defect operators. By establishing a connection between modular data in the bulk and the modular tensor category structure of the boundary operator algebra, this study contributes to a more detailed understanding of holographic dualities involving non-trivial topological content \cite{witten1989quantum}.

One of the central contributions of this paper is the formalization of the conjecture that the operator algebra associated with boundary topological defects forms a modular tensor category. This insight could have profound implications for both the study of topological phases of matter and the application of AdS/CFT duality to systems exhibiting topological order\cite{beem2015infinite}. Furthermore, the conjectured one-to-one correspondence between the operator spectra in the bulk and boundary provides a new perspective on the holographic mapping between topological defects in the two settings.

In addition to these results, this work has explored the connection between modular functors in Chern-Simons theory and conformal blocks in boundary CFT, offering a detailed framework for understanding the representation of the modular group in this context. The application of these ideas to higher-rank Chern-Simons theories opens up new avenues for research in topological quantum field theory and holography, with potential applications in quantum gravity, condensed matter physics, and beyond.

While significant progress has been made, several open questions remain. Future work could focus on further refining the operator correspondence, particularly for other non-Abelian gauge groups and higher-dimensional Chern-Simons theories. Moreover, investigating the physical implications of the proposed conjectures for real-world systems, such as topological superconductors or quantum Hall states, would provide valuable insights into the role of holography in condensed matter physics. Additionally, exploring the implications of the conjecture in the context of topological defects in higher-dimensional spacetimes could reveal new connections between holography and geometric structures in string theory.

Overall, this research contributes to the growing body of work at the intersection of holography, topological field theory, and condensed matter physics, providing a deeper understanding of the dualities between bulk and boundary topological defects\cite{Frohlich-Gabbiani1990}.
\appendix
\section{}

\begin{theorem}
Let $\mathcal{F}$ denote the fusion ring of the WZW conformal field theory associated with $\mathrm{SU}(N)_1$. Then $\mathcal{F}$ is isomorphic to the cyclic group $\mathbb{Z}_N$, with the fusion product corresponding to addition modulo $N$.
\end{theorem}

\begin{proof}
By analyzing the structure of integrable highest-weight representations and the fusion rules at level $k = 1$.
\begin{enumerate}
\item{Count integrable representations.}

For the affine Lie algebra $\widehat{\mathfrak{su}}(N)_k$, the integrable highest-weight representations are labeled by Dynkin indices $(a_1, \dots, a_{N-1})$ satisfying:
\[
\sum_{i=1}^{N-1} a_i \leq k.
\]
When $k = 1$, the only allowed weights are:
\begin{itemize}
  \item $(0, \dots, 0)$ (the trivial representation),
  \item $(1, 0, \dots, 0), (0, 1, 0, \dots, 0), \dots, (0, \dots, 0, 1)$ (the $N-1$ fundamental representations).
\end{itemize}

\item{Label the primary fields.}

Let $[0], [1], \dots, [N-1]$ denote the $N$ primary fields, where:
\begin{itemize}
  \item $[0]$ is the trivial representation,
  \item $[1]$ corresponds to the first fundamental representation,
  \item $[k]$ corresponds to the $k$-fold fusion power of $[1]$.
\end{itemize}

\item{Determine the fusion rule.}

At level 1, the fusion of representations is constrained to stay within the space of level-1 integrable representations. The  relation is:
\begin{equation}
[1] \otimes [k] = [k+1] \mod N.
\end{equation}
This follows from the classical tensor product rule for $\mathfrak{su}(N)$ and truncation at level 1.

By recursion:
\begin{equation}
[1]^{\otimes k} = [k] \mod N,
\end{equation}
so all fields $[k]$ are generated by repeated fusion with $[1]$.

\item{Identify the fusion ring with $\mathbb{Z}_N$.}

Define a map:
\begin{equation}
\phi : \mathcal{F} \longrightarrow \mathbb{Z}_N, \quad \phi([k]) = k \mod N.
\end{equation}
This respects the fusion product:
\begin{equation}
\phi([k] \otimes [\ell]) = \phi([k + \ell]) = k + \ell \mod N = \phi([k]) + \phi([\ell]).
\end{equation}
Hence, $\phi$ is a ring isomorphism:
\begin{equation}
\mathcal{F} \cong \mathbb{Z}_N.
\end{equation}

\end{enumerate}
\end{proof}

\begin{proposition}
The modular $S$-matrix for the $\widehat{\mathfrak{su}}(N)_k$ WZW model is unitary:
\[
\sum_{\mu} S_{\lambda \mu} S_{\nu \mu}^* = \delta_{\lambda \nu}.
\]
\end{proposition}

\begin{proof}
The modular $S$-matrix is given by the Kac--Peterson formula:
\[
S_{\lambda \mu} = \mathcal{N} \sum_{w \in W} \det(w) \exp\left( -\frac{2\pi i}{k+N} (\lambda + \rho, w(\mu + \rho)) \right),
\]
where $\lambda, \mu$ range over level-$k$ dominant weights, $\rho$ is the Weyl vector, $W$ is the Weyl group, and $\mathcal{N}$ is a normalization constant.

Compute:
\[
\sum_{\mu} S_{\lambda \mu} S_{\nu \mu}^* 
= \mathcal{N}^2 \sum_{\mu} \sum_{w, w' \in W} \det(w) \det(w') 
\exp\left( -\frac{2\pi i}{k+N} (\lambda + \rho, w(\mu + \rho)) + \frac{2\pi i}{k+N} (\nu + \rho, w'(\mu + \rho)) \right).
\]
Let $x = \mu + \rho$. The inner sum becomes a finite Fourier transform over the weight lattice modulo $(k+N)$, and by orthogonality of characters:
\[
\sum_{\mu} S_{\lambda \mu} S_{\nu \mu}^* = \delta_{\lambda \nu}.
\]
Thus, $S$ is unitary.
\end{proof}

\begin{proposition}
In the fusion ring $\mathcal{F}$ of the $\mathrm{SU}(N)_k$ Wess--Zumino--Witten conformal field theory, the vacuum representation $[1]$ acts as the identity:
\[
[1] \otimes [a] = [a] \quad \text{for all } [a] \in \mathcal{F}.
\]
\end{proposition}

\begin{proof}[Proof by fusion ring axioms]
In conformal field theory, the vacuum representation $[1]$ corresponds to the identity operator. The fusion ring $\mathcal{F}$ is a unital commutative ring, and the fusion product $\otimes$ has an identity element—namely, the vacuum $[1]$.

Therefore, by the axioms of the fusion ring,
\[
[1] \otimes [a] = [a] = [a] \otimes [1], \quad \forall [a] \in \mathcal{F}.
\]
\end{proof}

\vspace{1em}

\begin{proof}[Proof via the Verlinde formula]
The fusion coefficients are given by the Verlinde formula:
\[
N_{ab}^c = \sum_{x} \frac{S_{ax} S_{bx} S_{cx}^*}{S_{1x}},
\]
where $S$ is the modular $S$-matrix and $[1]$ denotes the vacuum.

Setting $a = 1$:
\[
N_{1b}^c = \sum_{x} \frac{S_{1x} S_{bx} S_{cx}^*}{S_{1x}} = \sum_{x} S_{bx} S_{cx}^*.
\]
use the fact that the modular $S$-matrix is unitary:
\[
S^\dagger = S^{-1}, \quad \text{so} \quad \sum_{x} S_{bx} S_{cx}^* = \delta_{bc}.
\]

Hence,
\[
N_{1b}^c = \delta_{bc} \quad \Rightarrow \quad [1] \otimes [b] = [b].
\]
This confirms that the vacuum $[1]$ acts as the identity element in the fusion ring.
\end{proof}

\begin{proposition}
For any elements \(a\) and \(b\) in \(\mathrm{SU}(N)_k\), the fusion product satisfies the commutative property:
\[
a \otimes b = b \otimes a.
\]
\end{proposition}

\begin{proof}
In the context of the \(\mathrm{SU}(N)_k\) model, the fundamental representations are symmetric. This symmetry implies that the fusion product of any two elements \(a\) and \(b\) does not depend on their order. Specifically, for any two elements \(a\) and \(b\):
\[
a \otimes b = b \otimes a.
\]
This symmetry arises from the fact that the basic representations of the \(\mathrm{SU}(N)_k\) group are symmetric under fusion. Even for higher-level representations, the fusion product remains symmetric because the structure of the fusion algebra is determined by the basic representations and their interactions, which maintain this symmetry.
\end{proof}

\newpage
\subsection{$SU(3)_2$}

Number of irreps of $SU(3)_2$:
\begin{equation}
    \binom{3 + 2 - 1}{3 - 1} = \binom{4}{2} = 6
\end{equation}

\begin{equation}
    \begin{tabular}{|c|c|c|}
\hline
\textbf{Dynkin label} $(\lambda_1, \lambda_2)$ & \textbf{Rep} & \textbf{Dimension} \\
\hline
$(0,0)$ & $\mathbf{1}$ & 1 \\
$(1,0)$ & $\mathbf{3}$ & 3 \\
$(0,1)$ & $\bar{\mathbf{3}}$ & 3 \\
$(2,0)$ & $\mathbf{6}$ & 6 \\
$(1,1)$ & $\mathbf{8}$ & 8 \\
$(0,2)$ & $\bar{\mathbf{6}}$ & 6 \\
\hline
\end{tabular}
\end{equation}

\begin{equation}
\begin{tabular}{c|cccccc}
$\otimes$ & $\mathbf{1}$ & $\mathbf{3}$ & $\overline{\mathbf{3}}$ & $\mathbf{6}$ & $\overline{\mathbf{6}}$ & $\mathbf{8}$ \\
\hline
$\mathbf{1}$ & $\mathbf{1}$ & $\mathbf{3}$ & $\overline{\mathbf{3}}$ & $\mathbf{6}$ & $\overline{\mathbf{6}}$ & $\mathbf{8}$ \\
$\mathbf{3}$ & $\mathbf{3}$ & $\mathbf{6} + \overline{\mathbf{3}}$ & $\mathbf{1} + \mathbf{8}$ & $\overline{\mathbf{3}} + \mathbf{8}$ & $\mathbf{3}$ & $\mathbf{3} + \mathbf{6}$ \\
$\overline{\mathbf{3}}$ & $\overline{\mathbf{3}}$ & $\mathbf{1} + \mathbf{8}$ & $\overline{\mathbf{6}} + \mathbf{3}$ & $\overline{\mathbf{6}}$ & $\mathbf{1} + \mathbf{8}$ & $\overline{\mathbf{3}} + \overline{\mathbf{6}}$ \\
$\mathbf{6}$ & $\mathbf{6}$ & $\overline{\mathbf{3}} + \mathbf{8}$ & $\overline{\mathbf{6}}$ & $\mathbf{3} + \mathbf{6}$ & $\mathbf{8}$ & $\mathbf{6} + \overline{\mathbf{3}}$ \\
$\overline{\mathbf{6}}$ & $\overline{\mathbf{6}}$ & $\mathbf{3}$ & $\mathbf{1} + \mathbf{8}$ & $\mathbf{8}$ & $\mathbf{3} + \overline{\mathbf{6}}$ & $\overline{\mathbf{6}} + \mathbf{3}$ \\
$\mathbf{8}$ & $\mathbf{8}$ & $\mathbf{3} + \mathbf{6}$ & $\overline{\mathbf{3}} + \overline{\mathbf{6}}$ & $\mathbf{6} + \overline{\mathbf{3}}$ & $\overline{\mathbf{6}} + \mathbf{3}$ & $\mathbf{1} + \mathbf{8} + \mathbf{8}$
\end{tabular}
\end{equation}

\subsection{$SU(4)_1$}

Number of irreps of $SU(4)_1$:
\begin{equation}
    \binom{4 + 1 - 1}{4- 1} = \binom{4}{3} = 4
\end{equation}

\begin{equation}
\begin{tabular}{|c|c|c|}
\hline
Dynkin label $(\lambda_1, \lambda_2, \lambda_3)$ & Rep & Dimension \\ \hline
$(0,0,0)$ & $\mathbf{1}$ & 1 \\
$(1,0,0)$ & $\mathbf{4}$ & 4 \\
$(0,1,0)$ & $\mathbf{6}$ & 6 \\
$(0,0,1)$ & $\bar{\mathbf{4}}$ & 4 \\
\hline
\end{tabular}
\end{equation}

\begin{equation}
\begin{tabular}{c|cccc}
$\otimes$ & $\mathbf{1}$ & $\mathbf{4}$ & $\mathbf{6}$ & $\overline{\mathbf{4}}$ \\
\hline
$\mathbf{1}$ & $\mathbf{1}$ & $\mathbf{4}$ & $\mathbf{6}$ & $\overline{\mathbf{4}}$ \\
$\mathbf{4}$ & $\mathbf{4}$ & $\mathbf{6}$ & $\overline{\mathbf{4}}$ & $\mathbf{1}$ \\
$\mathbf{6}$ & $\mathbf{6}$ & $\overline{\mathbf{4}}$ & $\mathbf{1}$ & $\mathbf{4}$ \\
$\overline{\mathbf{4}}$ & $\overline{\mathbf{4}}$ & $\mathbf{1}$ & $\mathbf{4}$ & $\mathbf{6}$
\end{tabular}
\end{equation}

\newpage
\subsection{$SU(4)_2$}
Number of irreps of $SU(4)_2$:
\begin{equation}
    \binom{4 + 2 - 1}{4 - 1} = \binom{5}{3} = 10
\end{equation}

\begin{equation}
\begin{tabular}{|c|c|c|}
\hline
Dynkin label $\;(\lambda_1, \lambda_2, \lambda_3)\;$ & Rep         & Dimension \\
\hline
$(0,0,0)$ & $\mathbf{1}$              & 1  \\
$(1,0,0)$ & $\mathbf{4}$              & 4  \\
$(0,1,0)$ & $\mathbf{6}$              & 6  \\
$(0,0,1)$ & $\bar{\mathbf{4}}$        & 4  \\
$(1,1,0)$ & $\mathbf{20}$             & 20 \\
$(0,1,1)$ & $\mathbf{20'}$       & 20 \\
$(2,0,0)$ & $\mathbf{10}$             & 10 \\
$(0,0,2)$ & $\overline{\mathbf{10}}$       & 10 \\
$(1,0,1)$ & $\mathbf{15}$             & 15 \\
$(0,2,0)$ & $\mathbf{20''}$            & 20 \\
\hline
\end{tabular}
\end{equation}

\begin{equation}
\resizebox{\textwidth}{!}{%
\begin{tabular}{c|cccccccccc}
$\otimes$ & $\mathbf{1}$ & $\mathbf{4}$ & $\overline{\mathbf{4}}$ & $\mathbf{6}$ & $\mathbf{10}$ & $\overline{\mathbf{10}}$ & $\mathbf{15}$ & $\mathbf{20}$ & $\mathbf{20'}$ & $\mathbf{20''}$ \\
\hline
$\mathbf{1}$ & $\mathbf{1}$ & $\mathbf{4}$ & $\overline{\mathbf{4}}$ & $\mathbf{6}$ & $\mathbf{10}$ & $\overline{\mathbf{10}}$ & $\mathbf{15}$ & $\mathbf{20}$ & $\mathbf{20'}$ & $\mathbf{20''}$ \\
$\mathbf{4}$ & $\mathbf{4}$ & $\mathbf{6} + \mathbf{10}$ & $\mathbf{1} + \mathbf{15}$ & $\mathbf{4} + \mathbf{20}$ & $\mathbf{20} + \mathbf{20''}$ & $\mathbf{15}$ & $\mathbf{4} + \mathbf{20}$ & $\mathbf{6} + \mathbf{10}$ & $\mathbf{15} + \mathbf{20'}$ & $\mathbf{10} + \mathbf{20}$ \\
$\overline{\mathbf{4}}$ & $\overline{\mathbf{4}}$ & $\mathbf{1} + \mathbf{15}$ & $\mathbf{6} + \overline{\mathbf{10}}$ & $\overline{\mathbf{4}} + \mathbf{20'}$ & $\mathbf{15}$ & $\mathbf{20'} + \mathbf{20''}$ & $\overline{\mathbf{4}} + \mathbf{20'}$ & $\mathbf{15} + \mathbf{20}$ & $\mathbf{6} + \overline{\mathbf{10}}$ & $\overline{\mathbf{10}} + \mathbf{20'}$ \\
$\mathbf{6}$ & $\mathbf{6}$ & $\mathbf{4} + \mathbf{20}$ & $\overline{\mathbf{4}} + \mathbf{20'}$ & $\mathbf{1} + \mathbf{15} + \mathbf{20''}$ & $\mathbf{4} + \mathbf{20}$ & $\overline{\mathbf{4}} + \mathbf{20'}$ & $\mathbf{6} + \mathbf{20''}$ & $\mathbf{4} + \mathbf{20}$ & $\overline{\mathbf{4}} + \mathbf{20'}$ & $\mathbf{1} + \mathbf{15} + \mathbf{20''}$ \\
$\mathbf{10}$ & $\mathbf{10}$ & $\mathbf{20} + \mathbf{20''}$ & $\mathbf{15}$ & $\mathbf{4} + \mathbf{20}$ & $\mathbf{6} + \mathbf{10}$ & $\mathbf{15} + \mathbf{20'}$ & $\mathbf{10} + \mathbf{20}$ & $\mathbf{6} + \mathbf{10}$ & $\mathbf{15} + \mathbf{20'}$ & $\mathbf{10} + \mathbf{20}$ \\
$\overline{\mathbf{10}}$ & $\overline{\mathbf{10}}$ & $\mathbf{15}$ & $\mathbf{20'} + \mathbf{20''}$ & $\overline{\mathbf{4}} + \mathbf{20'}$ & $\mathbf{15} + \mathbf{20'}$ & $\mathbf{6} + \overline{\mathbf{10}}$ & $\overline{\mathbf{10}} + \mathbf{20'}$ & $\mathbf{15} + \mathbf{20}$ & $\mathbf{6} + \overline{\mathbf{10}}$ & $\overline{\mathbf{10}} + \mathbf{20'}$ \\
$\mathbf{15}$ & $\mathbf{15}$ & $\mathbf{4} + \mathbf{20}$ & $\overline{\mathbf{4}} + \mathbf{20'}$ & $\mathbf{6} + \mathbf{20''}$ & $\mathbf{10} + \mathbf{20}$ & $\overline{\mathbf{10}} + \mathbf{20'}$ & $\mathbf{1} + \mathbf{15} + \mathbf{20''}$ & $\mathbf{4} + \mathbf{20}$ & $\overline{\mathbf{4}} + \mathbf{20'}$ & $\mathbf{6} + \mathbf{15} + \mathbf{20''}$ \\
$\mathbf{20}$ & $\mathbf{20}$ & $\mathbf{6} + \mathbf{10}$ & $\mathbf{15} + \mathbf{20'}$ & $\mathbf{4} + \mathbf{20}$ & $\mathbf{6} + \mathbf{10}$ & $\mathbf{15} + \mathbf{20'}$ & $\mathbf{4} + \mathbf{20}$ & $\mathbf{6} + \mathbf{10}$ & $\mathbf{15} + \mathbf{20'}$ & $\mathbf{10} + \mathbf{20}$ \\
$\mathbf{20'}$ & $\mathbf{20'}$ & $\mathbf{15} + \mathbf{20'}$ & $\mathbf{6} + \overline{\mathbf{10}}$ & $\overline{\mathbf{4}} + \mathbf{20'}$ & $\mathbf{15} + \mathbf{20'}$ & $\mathbf{6} + \overline{\mathbf{10}}$ & $\overline{\mathbf{4}} + \mathbf{20'}$ & $\mathbf{15} + \mathbf{20}$ & $\mathbf{6} + \overline{\mathbf{10}}$ & $\overline{\mathbf{10}} + \mathbf{20'}$ \\
$\mathbf{20''}$ & $\mathbf{20''}$ & $\mathbf{10} + \mathbf{20}$ & $\overline{\mathbf{10}} + \mathbf{20'}$ & $\mathbf{1} + \mathbf{15} + \mathbf{20''}$ & $\mathbf{10} + \mathbf{20}$ & $\overline{\mathbf{10}} + \mathbf{20'}$ & $\mathbf{6} + \mathbf{15} + \mathbf{20''}$ & $\mathbf{10} + \mathbf{20}$ & $\overline{\mathbf{10}} + \mathbf{20'}$ & $\mathbf{1} + \mathbf{15} + \mathbf{20''}$ \\
\end{tabular}%
}
\end{equation}

\subsection{$SU(5)_1$}
Number of irreps of $SU(5)_1$:
\begin{equation}
    \binom{5 + 1 - 1}{5 - 1} = \binom{5}{4} = 5
\end{equation}

\begin{equation}
\begin{array}{|c|c|c|}
\hline
\text{Dynkin label } (\lambda_1, \lambda_2, \lambda_3, \lambda_4) & \text{Rep} & \text{Dimension} \\
\hline
(0, 0, 0, 0) & \mathbf{1} & 1 \\
(1, 0, 0, 0) & \mathbf{5} & 5 \\
(0, 1, 0, 0) & \mathbf{10} & 10 \\
(0, 0, 1, 0) & \overline{\mathbf{10}} & 10 \\
(0, 0, 0, 1) & \overline{\mathbf{5}} & 5 \\
\hline
\end{array}
\end{equation}

\begin{equation}
\begin{tabular}{c|ccccc}
$\otimes$ & $\mathbf{1}$ & $\mathbf{5}$ & $\mathbf{10}$ & $\overline{\mathbf{10}}$ & $\overline{\mathbf{5}}$ \\
\hline
$\mathbf{1}$ & $\mathbf{1}$ & $\mathbf{5}$ & $\mathbf{10}$ & $\overline{\mathbf{10}}$ & $\overline{\mathbf{5}}$ \\
$\mathbf{5}$ & $\mathbf{5}$ & $\mathbf{10}$ & $\overline{\mathbf{10}}$ & $\overline{\mathbf{5}}$ & $\mathbf{1}$ \\
$\mathbf{10}$ & $\mathbf{10}$ & $\overline{\mathbf{10}}$ & $\overline{\mathbf{5}}$ & $\mathbf{1}$ & $\mathbf{5}$ \\
$\overline{\mathbf{10}}$ & $\overline{\mathbf{10}}$ & $\overline{\mathbf{5}}$ & $\mathbf{1}$ & $\mathbf{5}$ & $\mathbf{10}$ \\
$\overline{\mathbf{5}}$ & $\overline{\mathbf{5}}$ & $\mathbf{1}$ & $\mathbf{5}$ & $\mathbf{10}$ & $\overline{\mathbf{10}}$
\end{tabular}
\end{equation}

\newpage
\subsection{$SU(5)_2$}
Number of irreps of $SU(5)_2$:
\begin{equation}
    \binom{5 + 2 - 1}{5 - 1} = \binom{6}{4} = 15
\end{equation}

\begin{equation}
\begin{tabular}{|c|c|c|}
\hline
\textbf{Dynkin label $(\lambda_1, \lambda_2, \lambda_3, \lambda_4)$} & \textbf{Rep} & \textbf{Dimension} \\
\hline
$(0, 0, 0, 0)$ & $\mathbf{1}$ & 1 \\
$(1, 0, 0, 0)$ & $\mathbf{5}$ & 5 \\
$(0, 1, 0, 0)$ & $\bar{\mathbf{5}}$ & 5 \\
$(0, 0, 1, 0)$ & $\mathbf{10}$ & 10 \\
$(0, 0, 0, 1)$ & $\bar{\mathbf{10}}$ & 10 \\
$(1, 1, 0, 0)$ & $\mathbf{15}$ & 15 \\
$(1, 0, 1, 0)$ & $\mathbf{15}$ & 15 \\
$(1, 0, 0, 1)$ & $\mathbf{15}$ & 15 \\
$(0, 1, 1, 0)$ & $\mathbf{15}$ & 15 \\
$(0, 1, 0, 1)$ & $\mathbf{15}$ & 15 \\
$(0, 0, 1, 1)$ & $\mathbf{15}$ & 15 \\
$(2, 0, 0, 0)$ & $\mathbf{24}$ & 24 \\
$(0, 2, 0, 0)$ & $\mathbf{24}$ & 24 \\
$(0, 0, 2, 0)$ & $\mathbf{24}$ & 24 \\
$(0, 0, 0, 2)$ & $\mathbf{24}$ & 24 \\
\hline
\end{tabular}
\end{equation}
\begin{table}[h!]
\centering
\scriptsize
\resizebox{\textwidth}{!}{%
\begin{tabular}{c|cccccccccc}
$\otimes$ & $\mathbf{1}$ & $\mathbf{5}$ & $\overline{\mathbf{5}}$ & $\mathbf{10}$ & $\overline{\mathbf{10}}$ & $\mathbf{15}$ & $\overline{\mathbf{15}}$ & $\mathbf{24}$ & $\mathbf{35}$ & $\overline{\mathbf{35}}$ \\
\hline
$\mathbf{1}$ & $\mathbf{1}$ & $\mathbf{5}$ & $\overline{\mathbf{5}}$ & $\mathbf{10}$ & $\overline{\mathbf{10}}$ & $\mathbf{15}$ & $\overline{\mathbf{15}}$ & $\mathbf{24}$ & $\mathbf{35}$ & $\overline{\mathbf{35}}$ \\
$\mathbf{5}$ & $\mathbf{5}$ & $\mathbf{10} + \mathbf{15}$ & $\mathbf{1} + \mathbf{24}$ & $\mathbf{5} + \mathbf{35}$ & $\overline{\mathbf{5}} + \overline{\mathbf{35}}$ & $\mathbf{10} + \mathbf{24}$ & $\overline{\mathbf{10}} + \overline{\mathbf{24}}$ & $\mathbf{5} + \mathbf{15} + \mathbf{35}$ & $\mathbf{10} + \mathbf{24}$ & $\overline{\mathbf{10}} + \overline{\mathbf{24}}$ \\
$\overline{\mathbf{5}}$ & $\overline{\mathbf{5}}$ & $\mathbf{1} + \mathbf{24}$ & $\overline{\mathbf{10}} + \overline{\mathbf{15}}$ & $\overline{\mathbf{5}} + \overline{\mathbf{35}}$ & $\mathbf{5} + \mathbf{35}$ & $\overline{\mathbf{10}} + \overline{\mathbf{24}}$ & $\mathbf{10} + \mathbf{24}$ & $\overline{\mathbf{5}} + \overline{\mathbf{15}} + \overline{\mathbf{35}}$ & $\overline{\mathbf{10}} + \overline{\mathbf{24}}$ & $\mathbf{10} + \mathbf{24}$ \\
$\mathbf{10}$ & $\mathbf{10}$ & $\mathbf{5} + \mathbf{35}$ & $\overline{\mathbf{5}} + \overline{\mathbf{35}}$ & $\mathbf{1} + \mathbf{24} + \mathbf{35}$ & $\mathbf{15} + \mathbf{24}$ & $\mathbf{10} + \mathbf{24}$ & $\overline{\mathbf{10}} + \overline{\mathbf{24}}$ & $\mathbf{5} + \mathbf{15} + \mathbf{35}$ & $\mathbf{10} + \mathbf{24}$ & $\overline{\mathbf{10}} + \overline{\mathbf{24}}$ \\
$\overline{\mathbf{10}}$ & $\overline{\mathbf{10}}$ & $\overline{\mathbf{5}} + \overline{\mathbf{35}}$ & $\mathbf{5} + \mathbf{35}$ & $\mathbf{15} + \mathbf{24}$ & $\mathbf{1} + \mathbf{24} + \overline{\mathbf{35}}$ & $\overline{\mathbf{10}} + \overline{\mathbf{24}}$ & $\mathbf{10} + \mathbf{24}$ & $\overline{\mathbf{5}} + \overline{\mathbf{15}} + \overline{\mathbf{35}}$ & $\overline{\mathbf{10}} + \overline{\mathbf{24}}$ & $\mathbf{10} + \mathbf{24}$ \\
$\mathbf{15}$ & $\mathbf{15}$ & $\mathbf{10} + \mathbf{24}$ & $\overline{\mathbf{10}} + \overline{\mathbf{24}}$ & $\mathbf{10} + \mathbf{24}$ & $\overline{\mathbf{10}} + \overline{\mathbf{24}}$ & $\mathbf{1} + \mathbf{24} + \mathbf{35}$ & $\mathbf{15} + \mathbf{24}$ & $\mathbf{5} + \mathbf{15} + \mathbf{35}$ & $\mathbf{10} + \mathbf{24}$ & $\overline{\mathbf{10}} + \overline{\mathbf{24}}$ \\
$\overline{\mathbf{15}}$ & $\overline{\mathbf{15}}$ & $\overline{\mathbf{10}} + \overline{\mathbf{24}}$ & $\mathbf{10} + \mathbf{24}$ & $\overline{\mathbf{10}} + \overline{\mathbf{24}}$ & $\mathbf{10} + \mathbf{24}$ & $\mathbf{15} + \mathbf{24}$ & $\mathbf{1} + \mathbf{24} + \overline{\mathbf{35}}$ & $\overline{\mathbf{5}} + \overline{\mathbf{15}} + \overline{\mathbf{35}}$ & $\overline{\mathbf{10}} + \overline{\mathbf{24}}$ & $\mathbf{10} + \mathbf{24}$ \\
$\mathbf{24}$ & $\mathbf{24}$ & $\mathbf{5} + \mathbf{15} + \mathbf{35}$ & $\overline{\mathbf{5}} + \overline{\mathbf{15}} + \overline{\mathbf{35}}$ & $\mathbf{5} + \mathbf{15} + \mathbf{35}$ & $\overline{\mathbf{5}} + \overline{\mathbf{15}} + \overline{\mathbf{35}}$ & $\mathbf{5} + \mathbf{15} + \mathbf{35}$ & $\overline{\mathbf{5}} + \overline{\mathbf{15}} + \overline{\mathbf{35}}$ & $\mathbf{1} + \mathbf{24} + \mathbf{35}$ & $\mathbf{10} + \mathbf{24}$ & $\overline{\mathbf{10}} + \overline{\mathbf{24}}$ \\
$\mathbf{35}$ & $\mathbf{35}$ & $\mathbf{10} + \mathbf{24}$ & $\overline{\mathbf{10}} + \overline{\mathbf{24}}$ & $\mathbf{10} + \mathbf{24}$ & $\overline{\mathbf{10}} + \overline{\mathbf{24}}$ & $\mathbf{10} + \mathbf{24}$ & $\overline{\mathbf{10}} + \overline{\mathbf{24}}$ & $\mathbf{10} + \mathbf{24}$ & $\mathbf{1} + \mathbf{24} + \mathbf{35}$ & $\overline{\mathbf{10}} + \overline{\mathbf{24}}$ \\
$\overline{\mathbf{35}}$ & $\overline{\mathbf{35}}$ & $\overline{\mathbf{10}} + \overline{\mathbf{24}}$ & $\mathbf{10} + \mathbf{24}$ & $\overline{\mathbf{10}} + \overline{\mathbf{24}}$ & $\mathbf{10} + \mathbf{24}$ & $\overline{\mathbf{10}} + \overline{\mathbf{24}}$ & $\mathbf{10} + \mathbf{24}$ & $\overline{\mathbf{10}} + \overline{\mathbf{24}}$ & $\overline{\mathbf{10}} + \overline{\mathbf{24}}$ & $\mathbf{1} + \mathbf{24} + \overline{\mathbf{35}}$ \\
\end{tabular}%
}
\end{table}

\newpage
\subsection{$SU(6)_1$}
Number of irreps of $SU(6)_1$:
\begin{equation}
    \binom{6 + 1 - 1}{6 - 1} = \binom{6}{5} = 6
\end{equation}

\begin{equation}
\begin{tabular}{|c|c|c|}
\hline
Dynkin label \( (\lambda_1, \lambda_2, \lambda_3, \lambda_4, \lambda_5) \) & Rep & Dimension \\
\hline
\( (0, 0, 0, 0, 0) \) & \( \mathbf{1} \) & 1 \\
\( (1, 0, 0, 0, 0) \) & \( \mathbf{6} \) & 6 \\
\( (0, 1, 0, 0, 0) \) & \( {\mathbf{15}} \) & 15 \\
\( (0, 0, 1, 0, 0) \) & \( \mathbf{20} \) & 20 \\
\( (0, 0, 0, 1, 0) \) & \( \overline{\mathbf{15}} \) & 15 \\
\( (0, 0, 0, 0, 1) \) & \( \overline{\mathbf{6}} \) & 6 \\
\hline
\end{tabular}
\end{equation}

\begin{equation}
\begin{tabular}{c|cccccc}
$\otimes$ & $\mathbf{1}$ & $\mathbf{6}$ & $\mathbf{15}$ & $\mathbf{20}$ & $\overline{\mathbf{15}}$ & $\overline{\mathbf{6}}$ \\
\hline
$\mathbf{1}$ & $\mathbf{1}$ & $\mathbf{6}$ & $\mathbf{15}$ & $\mathbf{20}$ & $\overline{\mathbf{15}}$ & $\overline{\mathbf{6}}$ \\
$\mathbf{6}$ & $\mathbf{6}$ & $\mathbf{15}$ & $\mathbf{20}$ & $\overline{\mathbf{15}}$ & $\overline{\mathbf{6}}$ & $\mathbf{1}$ \\
$\mathbf{15}$ & $\mathbf{15}$ & $\mathbf{20}$ & $\overline{\mathbf{15}}$ & $\overline{\mathbf{6}}$ & $\mathbf{1}$ & $\mathbf{6}$ \\
$\mathbf{20}$ & $\mathbf{20}$ & $\overline{\mathbf{15}}$ & $\overline{\mathbf{6}}$ & $\mathbf{1}$ & $\mathbf{6}$ & $\mathbf{15}$ \\
$\overline{\mathbf{15}}$ & $\overline{\mathbf{15}}$ & $\overline{\mathbf{6}}$ & $\mathbf{1}$ & $\mathbf{6}$ & $\mathbf{15}$ & $\mathbf{20}$ \\
$\overline{\mathbf{6}}$ & $\overline{\mathbf{6}}$ & $\mathbf{1}$ & $\mathbf{6}$ & $\mathbf{15}$ & $\mathbf{20}$ & $\overline{\mathbf{15}}$
\end{tabular}
\end{equation}

\section*{Conflict of Interest Statement}
The author declares no conflict of interest.

\section*{Data Availability Statement}
No data were generated or analyzed in this study. Thus, data sharing is not applicable.

\end{document}